\documentclass[3p,times,twocolumn]{elsarticle}

\usepackage{ecrc}

\volume{00}

\firstpage{1}

\runauth{}

\jid{nuphbp}

\usepackage{amssymb}

\usepackage[figuresright]{rotating}

\begin{document}

\begin{frontmatter}




\vspace{-100 mm}

\title{Updated Results of a Solid-State Sensor Irradiation Study for ILC Extreme Forward Calorimetry \\ 
\bigskip Talk presented on behalf of the FCAL Collaboration at the International Workshop on Future Linear Colliders (LCWS2016), Morioka, 
   Japan, 5-9 December 2016. C16-12-05.4.}


\author{Paul Anderson}
\author{Wyatt Crockett}
\author{Luc D'Hauthuille}
\author{Vitaliy Fadeyev}
\author{Caleb Fink}
\author{Cesar Gonzalez Renteria}
\author{Benjamin Gruey}
\author{Jane Gunnell}
\author{Forest Martinez-McKinney}
\author{Greg Rischbieter}
\author{Kyle Rocha}
\author{Bruce A. Schumm \corref{cor1}}
\ead{baschumm@ucsc.edu}
\author{Edwin Spencer}
\author{Max Wilder}
\cortext[cor1]{Corresponding author.}
\address{Santa Cruz Institute for Particle Physics and the University Of California, 1156 High Street,
Santa Cruz California 95064 USA}

\begin{abstract}
Detectors proposed for the International Linear Collider (ILC)
incorporate a tungsten sampling calorimeter (`BeamCal') intended to
reconstruct showers of electrons, positrons and photons
that emerge from the interaction point of the collider
with angles between 5 and 50 milliradians. For the
innermost radius of this calorimeter, radiation doses
at shower max are expected to reach 100 Mrad per year,
primarily due to minimum-ionizing electrons and positrons
that arise in the induced electromagnetic showers
of e$^+$e$^-$ `beamstrahlung' pairs produced in the ILC beam-beam interaction. However,
radiation damage to calorimeter sensors may be dominated
by hadrons induced by nuclear interactions of shower photons,
which are much more likely to contribute to the non-ionizing
energy loss that has been observed to damage sensors exposed to
hadronic radiation. We report here on prior highlights and recent results of SLAC
Experiment T-506, for which several different types of
semiconductor sensors were exposed to doses of radiation
induced by showering electrons of energy 3.5-13.3 GeV. By embedding
the sensor under irradiation within a tungsten radiator, the exposure
incorporated hadronic species that would potentially contribute to the
degradation of a sensor mounted in a precision sampling calorimeter.
Depending on sensor technology, significant post-irradiation charge collection
was observed for doses of several hundred Mrad.

\end{abstract}


\end{frontmatter}


\section{Introduction}
\label{}

Far-forward calorimetry, covering the region between 5 and 50 milliradians
from the on-energy beam axis,
is envisioned as a component of both the ILD~\cite{ref:ILD_DBD} and SiD~\cite{ref:SiD_DBD}
detector concepts for the proposed International Linear Collider (ILC). The
BeamCal tungsten sampling calorimeter proposed to cover this angular region
is expected to absorb approximately 10 TeV of electromagnetic radiation
per beam crossing from e$^+$e$^-$ beamstrahlung pairs, leading to expected 
annual radiation doses of 100 Mrad
for the most heavily-irradiated portions of the instrument.
While the deposited energy is expected to arise primarily from minimum-ionizing
electrons and positrons in the induced electromagnetic showers,
radiation damage to calorimeter sensors may be dominated
by hadrons induced by nuclear interactions of shower photons,
which are much more likely to contribute to the non-ionizing
energy loss that has been observed to damage sensors exposed to
hadronic radiation. We report here on the latest results of SLAC
Experiment T-506, for which several different types of
solid-state sensors were exposed to doses of up to 300 Mrad
at the approximate maxima of electromagnetic
showers induced in a tungsten radiator by electrons of energy
3.5-13.3 GeV, similar to that of electrons and positrons
from ILC beamstrahlung pairs.

Bulk damage leading to the suppression of the electron/hole
charge-collection efficiency (CCE) is generally thought to be proportional
to the non-ionizing energy loss (`NIEL') component of the energy
deposited by the incident radiation.
Observations from early studies of electromagnetically-induced damage to
solar cells~\cite{ref:TaeSung5,ref:TaeSung6,ref:TaeSung7}
suggested that p-type bulk sensors were more tolerant to
damage from electromagnetic sources, due to an apparent
departure from NIEL scaling, particularly for electromagnetic particles
of lower incident energy.

Several more-recent studies have
explored radiation tolerance of silicon diode to incident fluxes of electrons.
A study assessing the capacitance vs. bias voltage (CV) characteristics of sensors exposed
to as much as 1 Grad of incident 2 MeV electrons~\cite{ref:Rafi_09}
suggested approximately 35 times less damage to n-type magnetic
Czochralski sensors than that expected from NIEL scaling.
A study of various n-type sensor types exposed to 900 MeV electrons
showed charge-collection loss of as little as 3\% for exposures up to
50 Mrad~\cite{ref:Dittongo2004}; for exposures of 150 Mrad, the degree of damage
was observed to be as small as one-fourth that expected from NIEL scaling~\cite{ref:Dittongo2005}. 
These discrepancies have been attributed to the different types of
defects created by lattice interactions: electrons tend to create point-like defects that are more
benign than the clusters formed due to hadronic interactions.

Finally, in studies of sensors exposed to large doses of hadron-induced
radiation, p-type bulk silicon was found to be more radiation-tolerant
than n-type bulk silicon, an observation that has been attributed to the absence of type inversion and the
collection of an electron-based signal~\cite{ref:TaeSung3,ref:TaeSung4}.
However, n-type bulk devices have certain advantages, such as a natural inter-electrode
isolation with commonly used passivation materials such as silicon oxide and silicon nitride.

More recently, a number of non-diode solid-state sensors have been proposed as a possible
radiation-tolerant alternative to silicon sensors, the latter of which can develop
a large dark current after significant irradiation.
Here, we report on an exploration of the radiation tolerance of silicon-diode pad sensors and
several bulk solid-state pad sensors, including gallium arsenide (GaAs), silicon carbide (SiC)
and industrial sapphire sensors. The sensors' radiation tolerance is
assessed via direct measurements of the median collected charge deposited
by minimum-ionizing particles as well as leakage current measurements.
Note that while diode sensors must be operated with the sense of bias voltage
that provides a reverse bias, bulk sensors can in principle be operated with
either sign of the bias voltage. When results are presented below for bulk sensors
with only one sign of the bias voltage, they are presented for the sign that provided the best
performance.

Two n-type and one p-type bulk silicon diode sensors were explored. The
p-bulk sensor (`WSI-P4' or `PF'), from a test structure associated 
with a prototype sensor developed for the ATLAS upgrade tracker~\cite{ref:nobu},
was fabricated by the Hamamatsu Photonics Corporation (HPK) from a float-zone crystal
with a thickness of 320 $\mu$m, and had an un-irradiated depletion voltage of 
approximately 180 V. The first of the n-bulk sensors (`N6906' or `NF'),
from a test structure associated with sensors developed for the 
Fermi satellite~\cite{ref:Ohsugi}, was also manufactured by HPK from
a float zone crystal, in this case with a thickness of 400 $\mu$m,
and an un-irradiated depletion voltage of less than 40 V.
In addition, the radiation tolerance of
a 320$\mu$m-thick n-type pad structure designed for use in the ILC Luminosity 
Calorimeter (LumiCal), which would cover the angular region just outside that of the BeamCal,
was explored. The sensor was manufactured by HPK; further
details about the sensor design can be found in~\cite{ref:Lumi}

This study made use of a fragment of a LumiCal sensor that had
been accidentally broken; once the broken sensor was obtained by SCIPP,
a study was done to determine, qualitatively, how the
leakage current of a given pad
depended upon that pad's proximity to the guard ring and the cleaved edge.
Based on this understanding, the fragment was intentionally cleaved into
two small pieces that could be mounted on the daughter boards used in the
radiation target and charge-collection-efficiency apparatus. After a regimen 
of exposing the sensor to elevated temperature, and exposing the cleaved 
edge to ultraviolet light, it was found that the sensor could be biased to over 
200 V without breaking down (i.e. developing a leakage current greater than several $\mu$A), well above the 
depletion voltage of 40 V. This allowed pre-irradiation charge-collection data
to be taken at full depletion. The irradiation process further raised the
breakdown voltage, allowing post-irradiation data to be taken at bias
voltages as high as 400 V.

The GaAs sensor used in the study was produced
by means of the Liquid Encapsulated Czochralski method doped by a shallow donor
(Sn or Te; Sn was used for the sensor in this study)~\cite{ref:GaAs},
and had a bulk thickness of 300 $\mu$m.
The SiC sensor used in the study featured double-layer nickel-gold
Schottky-barrier contacts mounted on a 4H-SiC crystal structure. The
sensors were 420 $\mu$m thick, with an high quality epitaxial (active)
layer of thickness 70 $\mu$m and an inactive substrate thickness of 350 $\mu$m.
More details about this sensor and its performance can be found in~\cite{ref:SiC}.
The industrial sapphire sensor used in the study had a thickness of 520 $\mu$m, 
and was produced by Crystal GmbH, Berlin. A three-layer metal contact 
consisting of aluminum (at the crystal surface) overlain with platinum 
and then gold was applied at the GSI laboratory in Darmstadt.

Prior results on the radiation tolerance of the GaAs sensor, as well as of
silicon diode sensors, including a p-type float-zone sensor irradiated to a 
dose of 270 Mrad, are presented in~\cite{ref:T506_NIM} 
and~\cite{ref:lcws2015}. New to this report are results for
n-type float-zone and sapphire sensors irradiated to 300 Mrad, and for
a 77 Mrad irradiation of the SiC sensor. Extended annealing studies
are also presented for the GaAs sensor and for the p-type float-zone sensor
that was irradiated to 270 Mrad.

While the radiation dose was initiated by electromagnetic processes
(electrons showering in tungsten), the placement of the sensors near
shower max ensures that the shower incorporates an appropriate
component of hadronic irradiation arising from neutron spallation,
photoproduction, and the excitation of the $\Delta$ resonance.
Particularly for the case that NIEL scaling suppresses
electromagnetically-induced radiation damage, the small
hadronic component of the electromagnetic
shower might dominate the rate of damage to the sensor.
However, the size and effect of this component is difficult to
estimate reliably, and so we choose to study radiation damage in
a configuration that naturally incorporates all components present in
an electromagnetic shower.

\section{Experimental Setup}

Un-irradiated sensors were subjected to current vs. bias voltage (IV) and 
capacitance vs. bias voltage (CV) tests,
the results of which allowed
a subset of them to be selected for irradiation based on their
breakdown voltage (typically above 1000 V for selected sensors) and low level of leakage
current. The sensors were placed on carrier printed-circuit `daughter boards' and wire-bonded to a
readout connector. The material of the daughter boards was milled away in the
region to be irradiated in order to facilitate the
charge collection measurement (described below) and minimize radio-activation.
The median collected charge was measured with the Santa Cruz Institute for Particle Physics (SCIPP)
charge-collection (CC) apparatus (also described below) before irradiation.
The sensors remained mounted to their individual daughter boards throughout irradiation and the followup
tests, simplifying their handling and reducing uncontrolled annealing.
Additionally, this allowed a reverse-bias voltage to be maintained across the sensor during irradiation.
This voltage was kept small (at the level of a few volts) to avoid possible damage of the devices
from a large instantaneous charge during the spill.

Sensors were irradiated with beam provided by the End Station
Test Beam (ESTB) facility at the SLAC National Accelerator Laboratory.
Parameters of the beam provided by the ESTB facility are shown in
Table~\ref{tab:ESTB}. The beam was incident upon a series of
tungsten radiators.
An initial 7mm-thick (2.0 radiation-length) tungsten plate (`R1')
served to initiate the electromagnetic shower.
The small number of
radiation lengths of this initial radiator permitted the
development of a small amount of divergence of the shower
relative to the straight-ahead beam direction without
significant development of the largely isotropic hadronic
component of the shower.

\begin{table}[h]
\begin{centering}
\caption{Parameters of the beam delivered by the
ESTB facility during the T-506 experiment.}
\label{tab:ESTB}
\vspace {5mm}
\begin{tabular}{cc}
Parameter  &  Value   \\ \hline
Energy     &  3.5-14.5 GeV \\
Repetition Rate  &  5-10 Hz \\
Charge per Pulse  &  150-180 pC \\
Spot Size (radius) & $\sim 1$ mm \\
\end{tabular}
\par\end{centering}
\end{table}

This plate was followed by an
open length of approximately half a meter, which allowed a degree of
spreading of the shower before it impinged upon a second,
significantly thicker `R2' (4.0 radiation-length) tungsten plate,
which was followed immediately by the sensor undergoing
irradiation. This was closely followed, in turn, by an
8.0 radiation-length tungsten plate. Immediately surrounding the
sensor by tungsten radiators that both
initiated and absorbed the great majority of the electromagnetic
shower ensured that the sensor would be illuminated by a
flux of hadrons commensurate with that experienced by a calorimeter
sensor close to the maximum of a tungsten-induced shower.
More precise values of the location of the various radiator elements
and sensor, for each of the four years of running of T-506, are given in Table~\ref{tab:radiator}.

\begin{table}[h]
\begin{centering}
\caption{Location of the various radiator elements and 
the sensor under irradiation, for the three successive 
phases of T-506 running. The R1 radiator had a thickness
of 2 $X_0$, while the thickness of the R2 radiator was 4 $X_0$.
The apparent increased geometrical thickness of R2 in Year 1
was due to the presence of a 6mm air gap mid-way through the
radiator.}
\label{tab:radiator}
\vspace {5mm}
\begin{tabular}{lccc}
                &  Year 1        & Year 2        & Year 3-4  \\
                &  (2013)        &  (2014)       &  (2015-6) \\
Surface         &  Location      & Location      & Location  \\
                &   (cm)         &    (cm)       & (cm)      \\
\hline
R1 Entrance  &  0.0  & 0.0  & 0.0  \\
R1 Exit      &  0.7  & 0.7  & 0.7  \\
R2 Entrance  &  55.0 & 45.7 & 46.6 \\
R2 Exit      &  57.0 & 47.1 & 48.0 \\
Sensor       &  57.7 & 47.6 & 48.5 \\
\end{tabular}
\par\end{centering}
\end{table}

Although initiating the shower significantly upstream of the sensor
promoted a more even illumination of the sensor
than would otherwise have been achieved, the half-width
of the resulting electron-positron fluence distribution
at the sensor plane was less than 0.5 cm. On the other hand,
the aperture of the CC apparatus (to be
described below) was of order 0.7 cm. Thus, in order to
ensure that the radiation dose was well understood over
the region of exposure to the CC apparatus source,
it was necessary to achieve a uniform illumination over
a region of approximately 1 cm$^2$. This was done by
`rastering' the detector across the beam spot through
a range of 1 cm in both dimensions
transverse to that of the incident beam.
According to Monte Carlo simulation studies, this is expected
to generate a region of approximately 1 cm$^2$ over which
the illumination is uniform to within $\pm 20$\%. To account
for potential millimeter-level misalginments of the beamline
center with the sensor, a `targeting factor' of $(90 \pm 10$)\%
is included in the final dose-rate calculations.

\section{Dose Rates}

During the 120 Hz operation of the SLAC Linac Coherent Light Source (LCLS),
5-10 Hz of beam was deflected by a pulsed kicker magnet into the End Station transfer line.
The LCLS beam was very stable with respect to both current and energy. Electronic
pickups and ionization chambers measured the beam current and beam loss through the
transfer line aperture, ensuring that good transfer efficiency could be established
and maintained. The transfer efficiency 
was estimated to be ($95 \pm 5$)\%. 

To calculate the dose rate through the sensor, it is necessary to determine
the `shower conversion factor' $\alpha$ that provides the mean fluence of minimum-ionizing
particles (predominantly electrons and positrons), in particles per cm$^2$,
per incoming beam electron. This factor is dependent upon the radiator
configuration and incident beam energy, as well as the rastering pattern
used to provide an even fluence across the sensor (as stated above,
the detector was translated continuously across the beam centerline
in a 1 cm$^2$ square pattern).

To estimate $\alpha$, the Electron-Gamma-Shower (EGS) Monte Carlo program~\cite{ref:EGS}
was used to simulate showers through the radiator configuration
and into the sensor. The radiator configuration
was input to the EGS program, and a mean fluence profile (particles per
cm$^2$ through the sensor as a function of transverse distance from the nominal
beam trajectory) was accumulated by simulating the showers of 1000
incident electrons of a given energy. To simulate the rastering process,
the center of the simulated profile was then
moved across the face of the sensor in 0.5mm steps, and an estimated mean fluence
per incident electron as a function of position on the sensor (again, relative to the nominal beam
trajectory) was calculated. This resulted in a mean fluence per incident electron
that was uniform to within (as stated above) $\pm$20\% anywhere inside the boundary of the rastering region.
The value of $\alpha$ used for subsequent irradiation dose estimates was taken to be
the value found at the intersection of the nominal
beam trajectory with the sensor plane. The simulation was repeated for
various values of the incident electron energy, producing the values of
$\alpha$ shown in Table~\ref{tab:alpha_2013} (Table~\ref{tab:alpha_2014-15})
for the 2013 (2014-16) radiator configuration.
For years 2014 through 2016, the spacings
of the radiator and sensor were similar enough that
a single mean value of $\alpha$ sufficed.

\begin{table}[h]
\begin{centering}
\caption{Shower conversion factor $\alpha$, giving
the mean fluence at the sensor per incident
electron, as a function of electron energy. for the
2013 radiator configuration. These
values include the effect of rastering over a 1 cm$^2$
area surrounding the nominal beam trajectory.
Also shown is the number of rads per nC of delivered
charge, at the given energy, corresponding to the
given value of $\alpha$. 
}
\label{tab:alpha_2013}
\vspace {5mm}
\begin{tabular}{ccc}
Beam      & 2013 Shower                     & Dose per nC             \\
Energy    &   Conversion                    & Delivered                \\
(GeV)     & Factor $\alpha$                 & Charge (krad)           \\ \hline
2  &  2.1 & 0.34  \\
4  &  9.4 & 1.50  \\
6  & 16.5 & 2.64  \\
8  & 23.5 & 3.76  \\
10 & 30.2 & 4.83  \\
12 & 36.8 & 5.89  \\
\end{tabular}
\par\end{centering}
\end{table}

\begin{table}[h]
\begin{centering}
\caption{Shower conversion factor $\alpha$, giving
the mean fluence at the sensor per incident
electron, as a function of electron energy, 
for the 2014-16 radiator configuration. These
values include the effect of rastering over a 1 cm$^2$
area surrounding the nominal beam trajectory.
Also shown is the number of rads per nC of delivered
charge, at the given energy, corresponding to the
given value of $\alpha$. For 2014 through 2016, the spacings
of the radiator and sensor were similar enough that
a single mean value of $\alpha$ sufficed.
}
\label{tab:alpha_2014-15}
\vspace {5mm}
\begin{tabular}{ccc}
Beam      & 2014-16 Shower      & Dose per nC             \\
Energy    & Conversion          & Delivered                \\
(GeV)     & Factor $\alpha$     & Charge (krad)           \\ \hline
3  & 4.6 &  0.73  \\
5  & 10.0 & 1.60  \\
7  & 15.5 & 2.48  \\
9  & 21.1 & 3.38  \\
11 & 26.7 & 4.27  \\
13 & 31.8 & 5.09  \\
15 & 37.7 & 6.03  \\
17 & 43.0 & 6.88  \\
\end{tabular}
\par\end{centering}
\end{table}

To convert this number to rads per nC of delivered charge, a mean
energy loss in silicon of 3.7 MeV/cm was assumed, leading to
a fluence-to-rad conversion factor of 160 rad per nC/cm$^2$.
It should be noted that while this dose rate considers only
the contribution from electrons and positrons, these
two sources dominate the overall energy absorbed by the
sensor. In addition, the BeamCal dose-rate spec of 100 Mrad
per year considered only the contribution from electrons
and positrons.

To confirm the adequacy of the dose-calibration simulation, 
in 2013 an
in-situ measurement of the dose was made using a
radiation-sensing field-effect transistor (`RADFET')~\cite{ref:radfet}
positioned on a daughter board at the expected
position of the nominal beam trajectory at the
center of the rastering pattern.
Beam was delivered in 150 pC pulses of 4.02 GeV
electrons; a total of 1160 pulses were directed
into the target over a period of four minutes,
during which the sensor was rastered quickly
through its 1 cm$^2$ pattern.
The RADFET was then read out, indicating
a total accumulated dose of 230 krad,
with an uncertainty of roughly 10\%. Making
use of the dose rate calibration of Table~\ref{tab:alpha_2013},
interpolating to the exact incident energy of 4.02 GeV,
and taking into account the ($95 \pm 5$)\% transfer efficiency
of the ESTB beamline, leads to an expected dose of 250 krad,
within the $\sim$10\% uncertainty of the RADFET measurement.

\section{Sensor Irradiation Levels}

This proceeding reports on the study of three types of silicon diode sensors.
Two pad sensors with p-type and n-type bulk doping (denoted ``PF'' and ``NF'', respectively) 
were produced from float-zone crystals. A third n-type silicon diode
sensor (denoted ``LUMI'') was designed for use in the ILC Luminosity Calorimeter.
In addition, irradiated bulk (non-diode) 
GaAs, SiC and industrial sapphire sensors were studied.
Once a sensor was irradiated in a $0^{\circ}-5^{\circ}$ C environment
at the ESTB, it was placed
in a sub-freezing environment and not irradiated again.
Up to four sensors of each type were irradiated and
chilled until they could be brought back to the University
of California, Santa Cruz campus for the
post-irradiation CC and leakage current measurements. In
addition, the sub-freezing environment was maintained
both during and after the CCE and current measurements, so
that controlled annealing studies could be performed.
Table~\ref{tab:dose} displays the dose parameters of the
irradiated sensors. The
$(95 \pm 5)$\% transfer line efficiency 
and the $(90 \pm 10)$\% targeting factor have been taken
into account in these estimates. 

\begin{table*}[h]
\begin{centering}
\caption{Dose parameters of the irradiated sensors. A
$(95 \pm 5)$\% transfer line efficiency and
a $(90 \pm 10)$\% targeting factor has been taken
into account in final dose estimates.}
\label{tab:dose}
\begin{tabular}{lcccc}
Sensor  & Year &  Beam Energy         &   Delivered      &  Dose     \\
        &      & (GeV)                &  Charge ($\mu$C) & (Mrad)    \\ \hline \hline
WSI-P4 (PF)     & 2015 &  13.3       &  50.9        & 269   \\           
N6906 (NF)      & 2015 &  14.6       &  60.0        & 290   \\   
LUMI            & 2016 & (14.5,13.4) & (30.3,37.0)  & 316   \\ \hline
GaAs-09         & 2014 &  3.90       &  21.7        & 20.8  \\ \hline
SiC             & 2015 &  13.3       &  17.2        & 76.6    \\ \hline
Sapphire-04     & 2016 & (14.5,13.4) & (13.1,53.8)  & 307   \\ \hline
\end{tabular}
\par\end{centering}
\end{table*}

\section{Charge Collection Measurement}

The SCIPP CC
apparatus incorporates a $^{90}$Sr source that has a secondary $\beta$-decay
with an end-point energy of 2.28 MeV that illuminate
the sensor under study, passing through to a scintillator immediately behind
the sensor that is read out by a photomultiplier tube.

For assessing the CCE of pad sensors, 
a two-stage, single-channel amplifier was constructed from discrete components,
based on a design of Fabris, Madden and Yaver~\cite{ref:jfet_amp}.
For the first stage, a cascode of two NXP BF862 JFETs is used. The source of the second 
JFET was connected to the non-inverting input of an LM6171 operational amplifier,
chosen for its high slew rate and low input noise contribution. 
The output of this opamp was then fed back to the input of the first JFET through
a $0.05$ pF capacitor shunted by a 10 M$\Omega$ resistor, 
completing the negative feedback loop. An external network, including 
a 32 dB Sonoma Instrument 310 SDI amplifier, was used to further amplify the pulse and 
shape it to a rise-time of 290 ns. 

\begin{figure}[h]
 \begin{center}
   \includegraphics[width=0.45\textwidth]{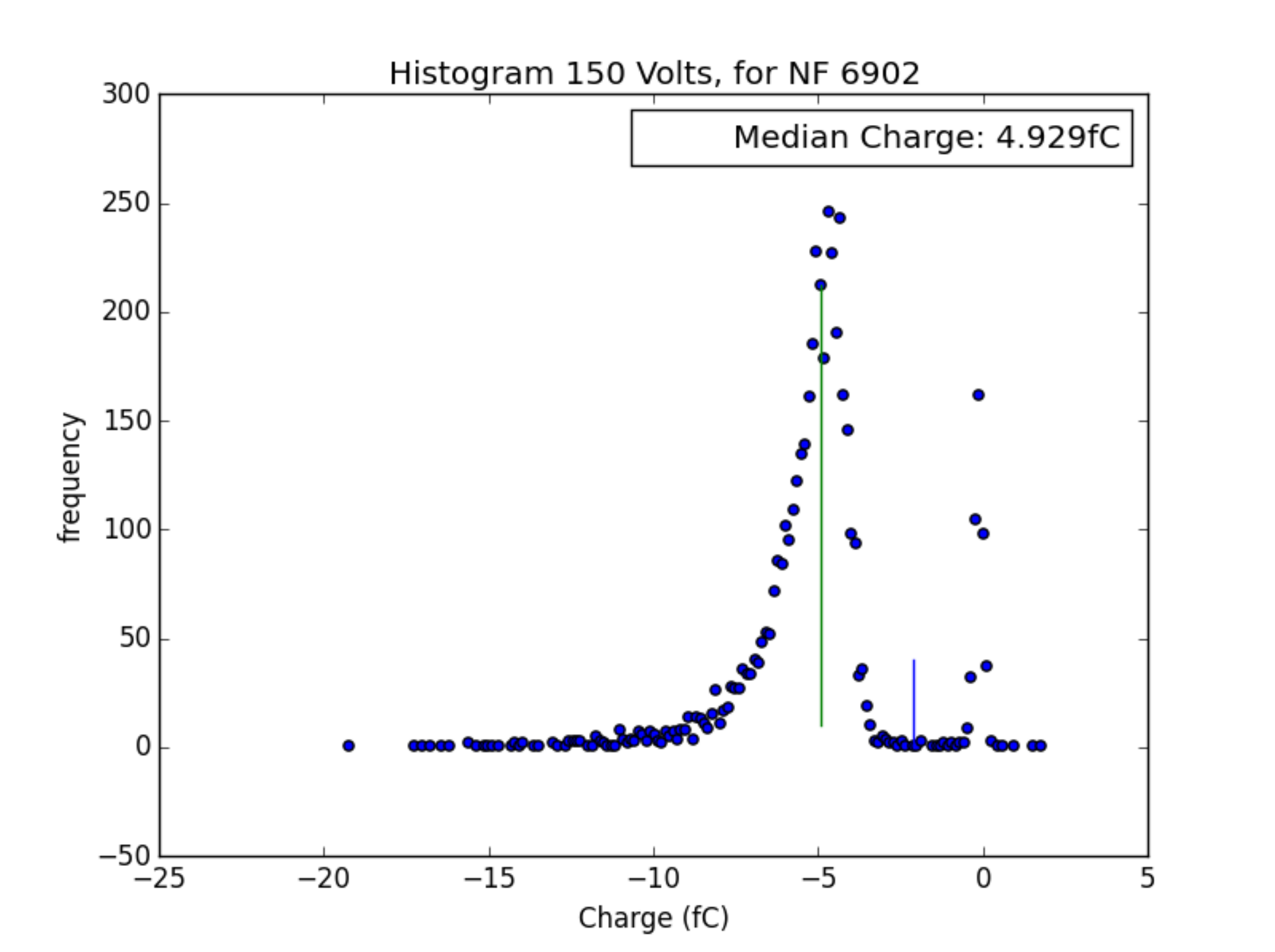}
 \end{center}
 \caption{
Histogram of pulse height for photomultiplier-triggered data events for
the single-channel readout. Both
the Landau distribution due to through-going $\beta$ particles as well
as the noise pedestal (for triggers for which the $\beta$ particle did not
traverse the sensor) are seen.
\label{fig:PH_dist}
}
\end{figure}

Upon receiving a trigger from the scintillator, the signal 
from the amplifier was read out by a Tektronix DPO 4054
digital storage oscilloscope, and the digitized waveforms
were written out and stored on the disk of a dedicated
data-acquisition computer. After the waveforms were accumulated on the computer,
a narrow temporal window was set around the peak of the average excitation pulse
from through-going beta particles, and a histogram was made of the resulting
pulse-height distribution; a typical distribution is shown in Figure~\ref{fig:PH_dist}.
Since not all $\beta$ particles that trigger the scintillator go through the pad,
the distribution shows contributions from both the Landau deposition of the
through-going $\beta$ particles, as well as that of the noise pedestal, allowing
for an in-situ subtraction of the mean pedestal.

The amplification system was calibrated by reading out an unirradiated silicon diode
sensor of known thickness, and
comparing the median charge of the resulting Landau
distribution (after subtracting off the mean pedestal) to that expected
for an unirradiated sensor of that thickness.
The measured gain exhibited only a small dependence on load
capacitance. The width of the pedestal distribution then provides a measurement of the 
readout noise, which was found to be approximately 250 electrons at room temperature. 

\section{Charge Collection and Leakage Current Results}

The daughter boards
containing the irradiated sensors were designed
with connectors that allowed them to be attached to the
CC apparatus readout board without handling the sensors.
The median CC was measured as a function of bias voltage for each sensor
both before and after irradiation, typically after a series of hour-long
annealing steps at successively higher temperatures.

\subsection{Results for bulk (non-diode) sensors}

\begin{figure}[h]
 \begin{center}
   \includegraphics[width=0.50\textwidth]{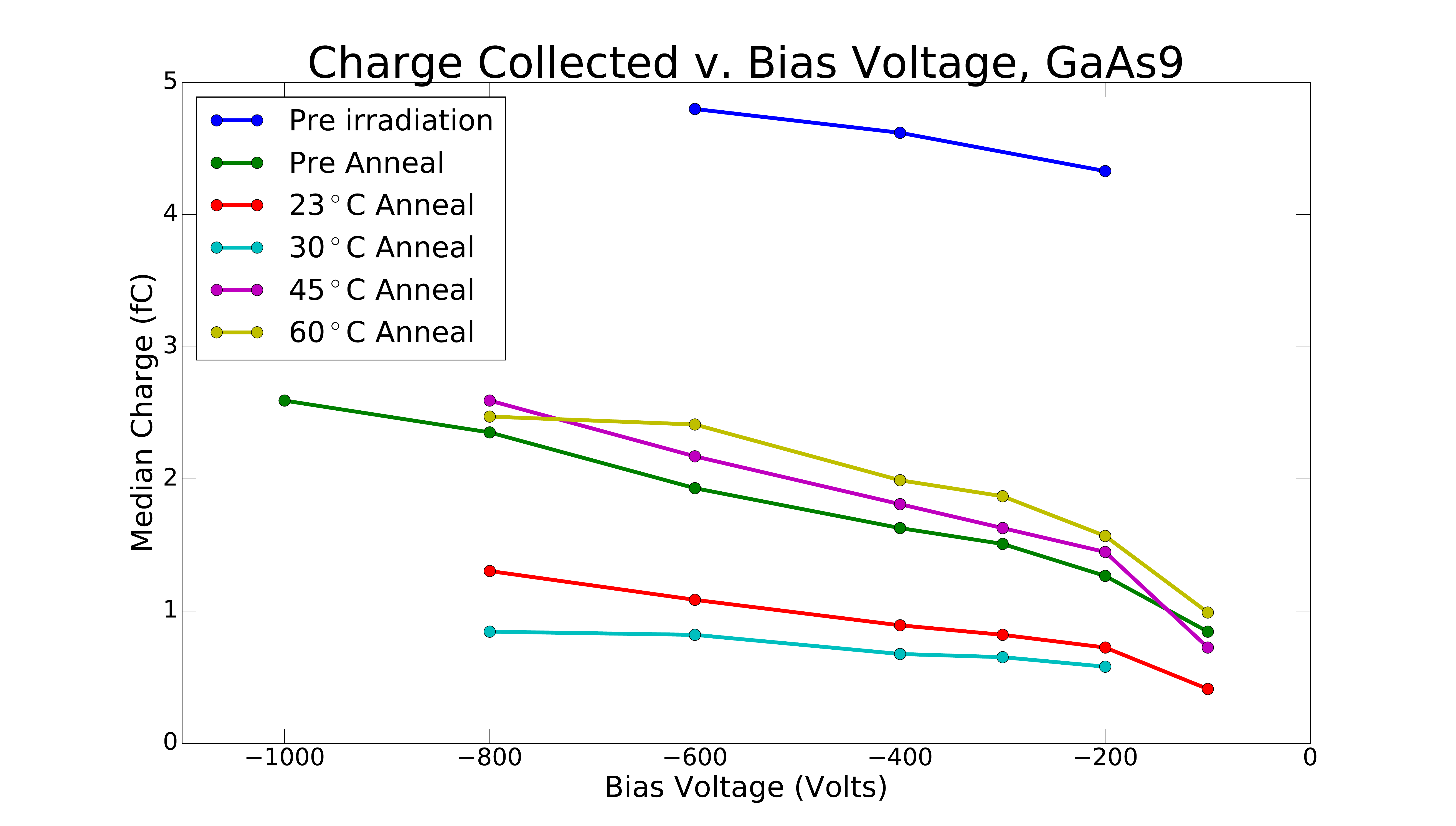}
 \end{center}
 \caption{
Dependence of the median collected charge from a GaAs sensor upon bias
voltage and annealing temperature, after exposure to a dose of 21 Mrad
of electromagnetically-induced radiation. Also shown is the median
collected charge as a function of bias voltage prior to irradiation. 
\label{fig:GaAs_traj}
}
\end{figure}

\begin{figure}[h]
 \begin{center}
   \includegraphics[width=0.50\textwidth]{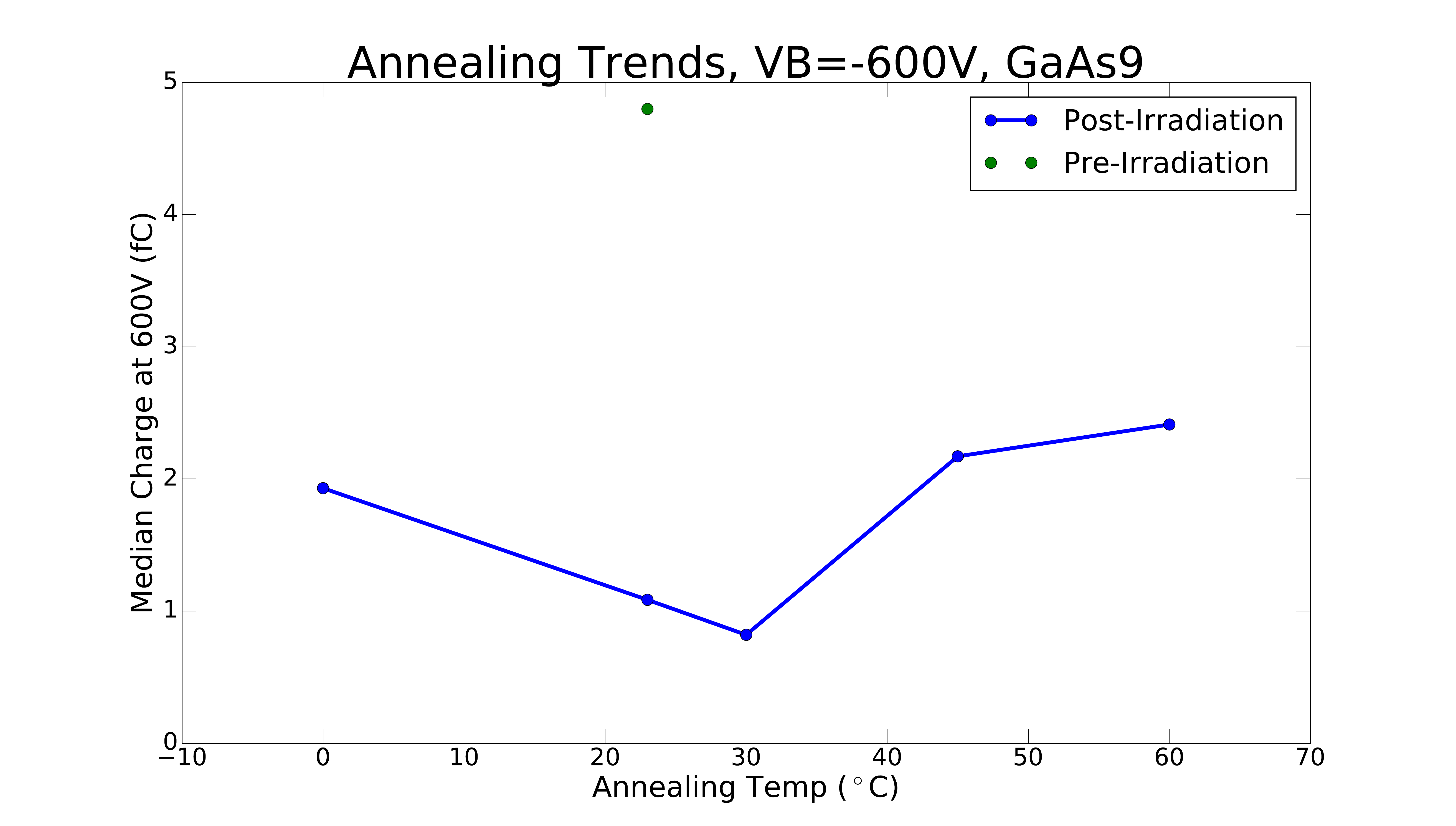}
 \end{center}
 \caption{
Dependence of the median collected charge from a GaAs sensor upon 
annealing temperature for a bias of $V_B$ = -600 V, before and after exposure to a dose of 21 Mrad
of electromagnetically-induced radiation.
\label{fig:GaAs_slice}
}
\end{figure}

\begin{figure}[h]
 \begin{center}
   \includegraphics[width=0.40\textwidth]{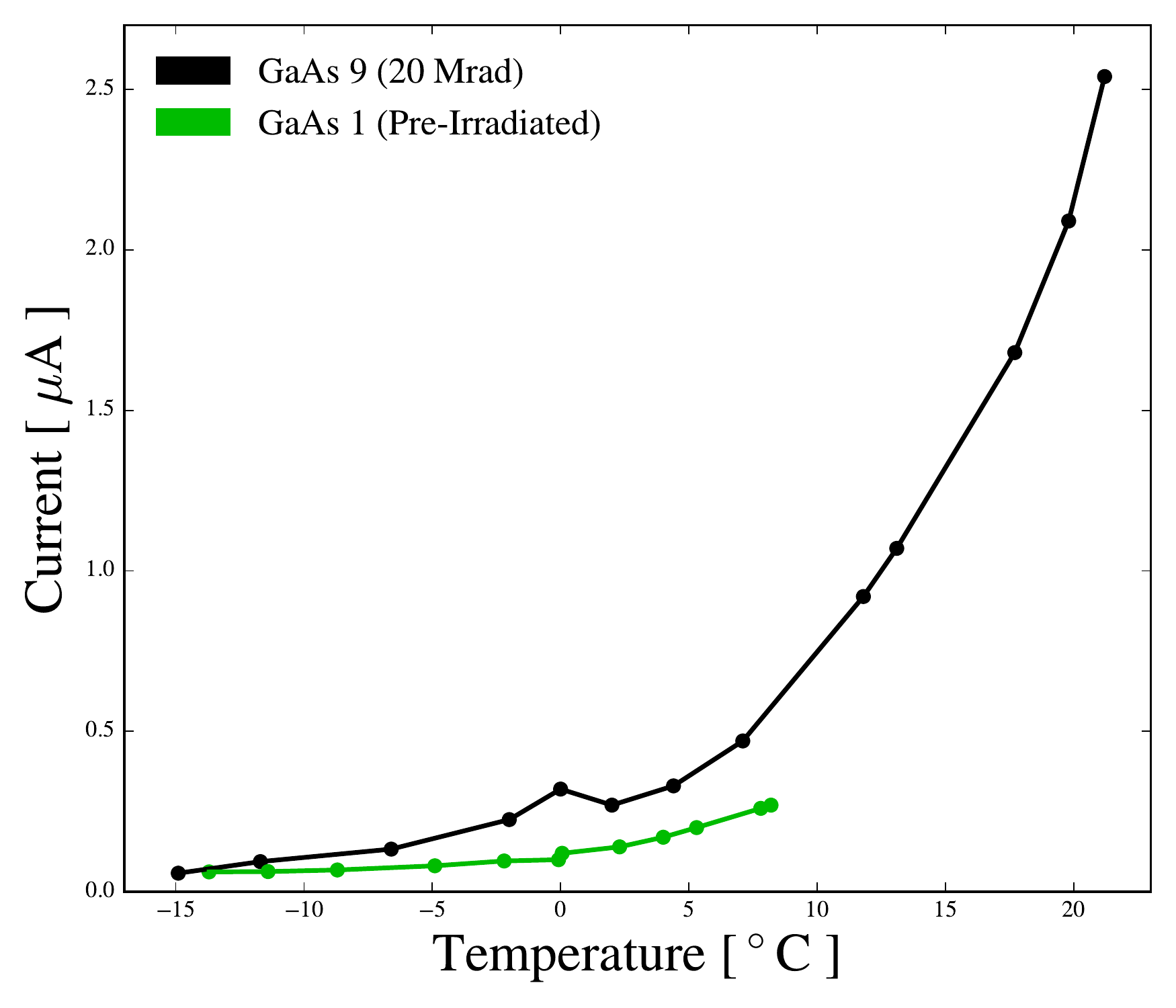}
 \end{center}
 \caption{
Leakage current vs. temperature for unirradiated and irradiated (21 Mrad)
GaAs sensors. The study was done with a bias of $V_B$ = -600 V. The irradiated GaAs sensor 
had been annealed for an hour at a temperature of 75$^o$ C.
\label{fig:GaAs_current}
}
\end{figure}

GaAs has been made use of in sensors designed specifically for use in the BeamCal
instrument. The GaAs test structure described in Section~1 was irradiated to a 
level of 21 Mrad in 2014. 
Figure~\ref{fig:GaAs_traj} exhibits the observed CCE for the
GaAs sensor as a function of bias voltage and annealing temperature. Figure~\ref{fig:GaAs_slice}
shows the CCE as a function of annealing temperature.
The sensor exhibited a significant loss in CCE, which
worsened after low-temperature annealing but then recovered somewhat after
higher-temperature annealing. Figure~\ref{fig:GaAs_current} shows the sensor's pre- and
post-irradiation leakage current as a function of temperature, for a bias
voltage of $V_B$ = -600 V. A significant dependence on 
sensor temperature is observed, as well as a degradation (increase) after irradiation.
This increased leakage current was not observed to improve significantly with annealing.

\begin{figure}[h]
 \begin{center}
   \includegraphics[width=0.50\textwidth]{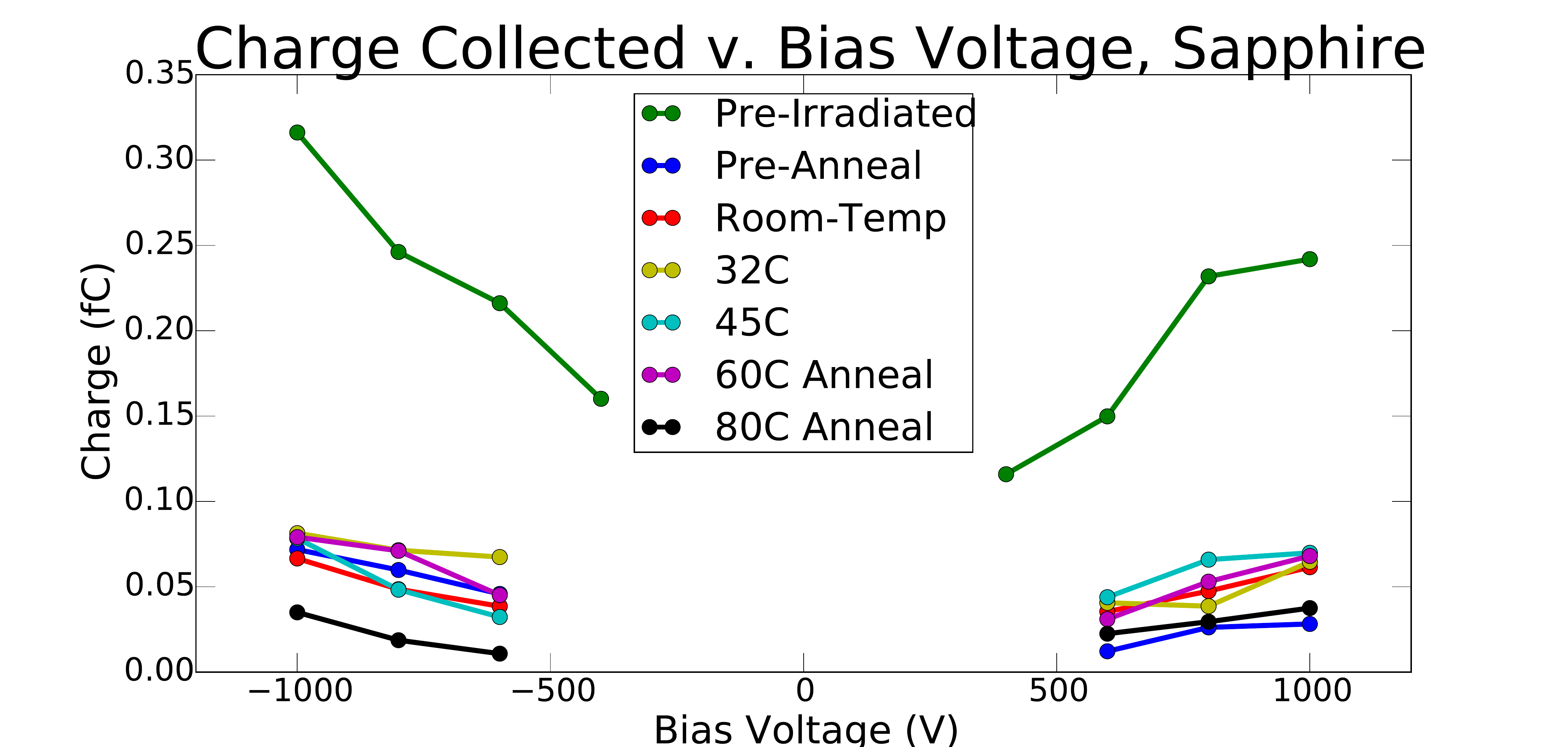}
 \end{center}
 \caption{
Dependence of the median collected charge from an industrial sapphire sensor upon bias
voltage and annealing temperature, after exposure to a dose of 307 Mrad
of electromagnetically-induced radiation. Also shown is the median
collected charge as a function of bias voltage prior to irradiation.
\label{fig:Sap_traj}
}
\end{figure}

\begin{figure}[h]
 \begin{center}
   \includegraphics[width=0.50\textwidth]{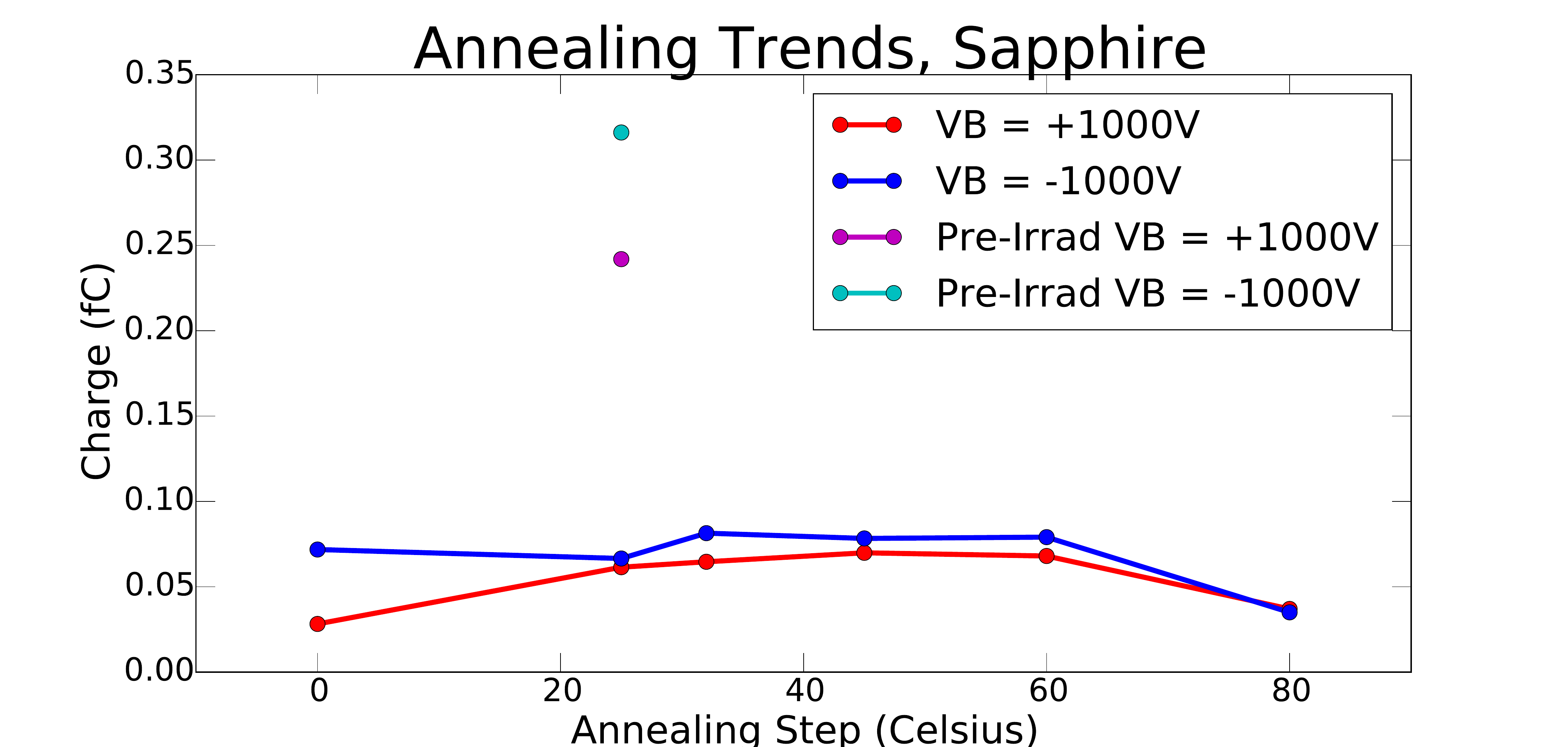}
 \end{center}
 \caption{
Dependence of the median collected charge from an industrial sapphire sensor upon
annealing temperature for biases of $\pm 1000$ V, before and after exposure to a dose of 307 Mrad
of electromagnetically-induced radiation.
\label{fig:Sap_slice}
}
\end{figure}

Industrial sapphire has been proposed as a possible sensor technology for the 
BeamCal, due to an expectation that it will exhibit radiation tolerance similarly favorable
to that of industrial diamond, which is much more costly than industrial sapphire.
The intrinsic CCE is low, however, presumably 
due to a short mean-free path of carriers in the sensor bulk. After exposure
to 307 Mrad of electromagnetically-induced radiation, leakage current remained
in the nanoamp range over the ~1 cm$^2$ area of the sensor. As illustrated in
Figure~\ref{fig:Sap_traj}, the median collected charge before irradiation
was measured to be only of order 0.3 fC for a detector basis as high as 1000 V,
with a significant drop in CCE observed after irradiation. Figure~\ref{fig:Sap_slice}
shows the measure CCE as a function of annealing temperature for biases
of $\pm 1000$ V; no improvement is observed over the course of the annealing process.

\begin{figure}[h]
 \begin{center}
   \includegraphics[width=0.50\textwidth]{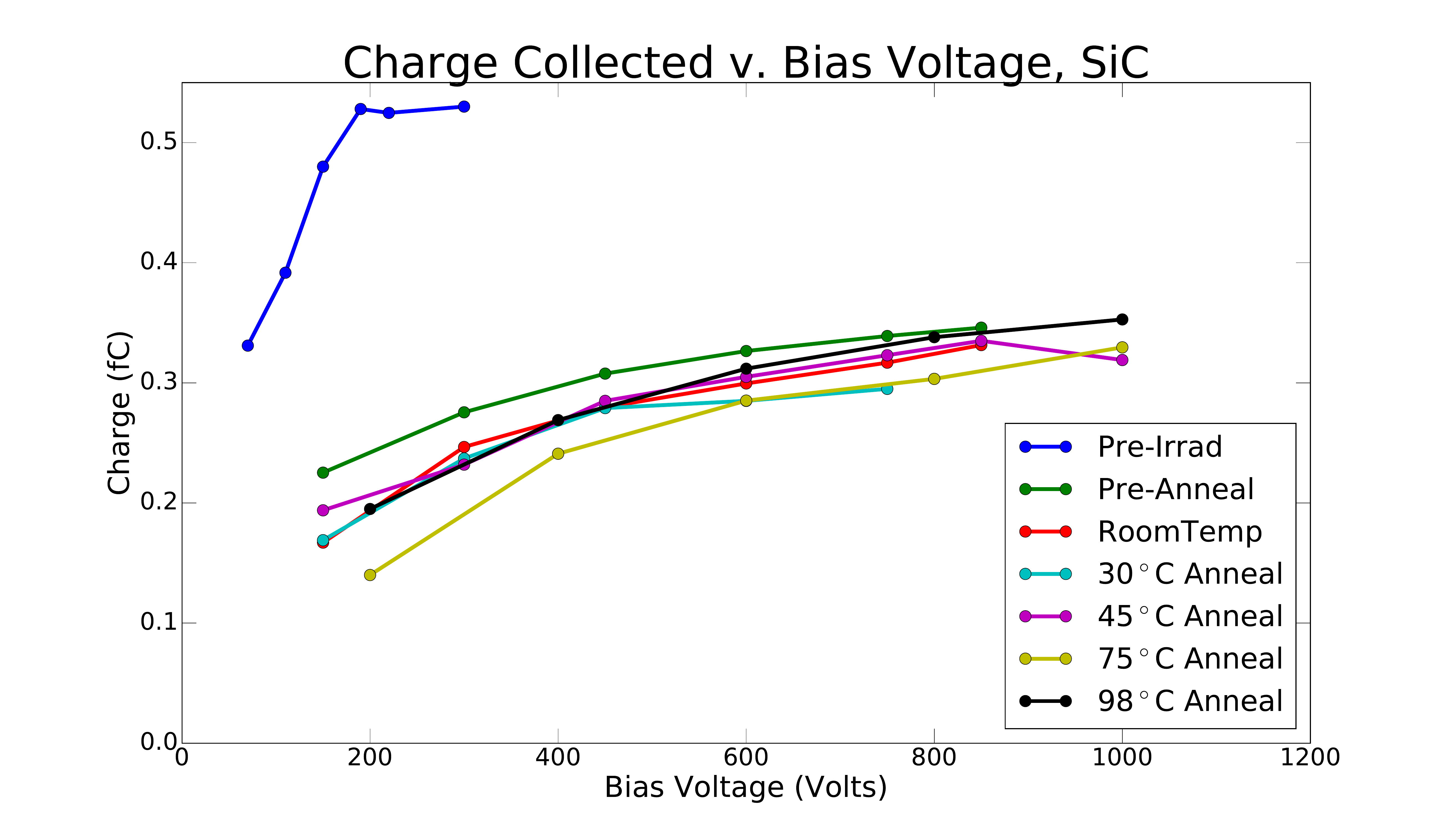}
 \end{center}
 \caption{
Dependence of the median collected charge from a SiC sensor upon bias
voltage and annealing temperature, after exposure to a dose of 77 Mrad
of electromagnetically-induced radiation. Also shown is the median
collected charge as a function of bias voltage prior to irradiation.
\label{fig:SiC_traj}
}
\end{figure}

\begin{figure}[h]
 \begin{center}
   \includegraphics[width=0.50\textwidth]{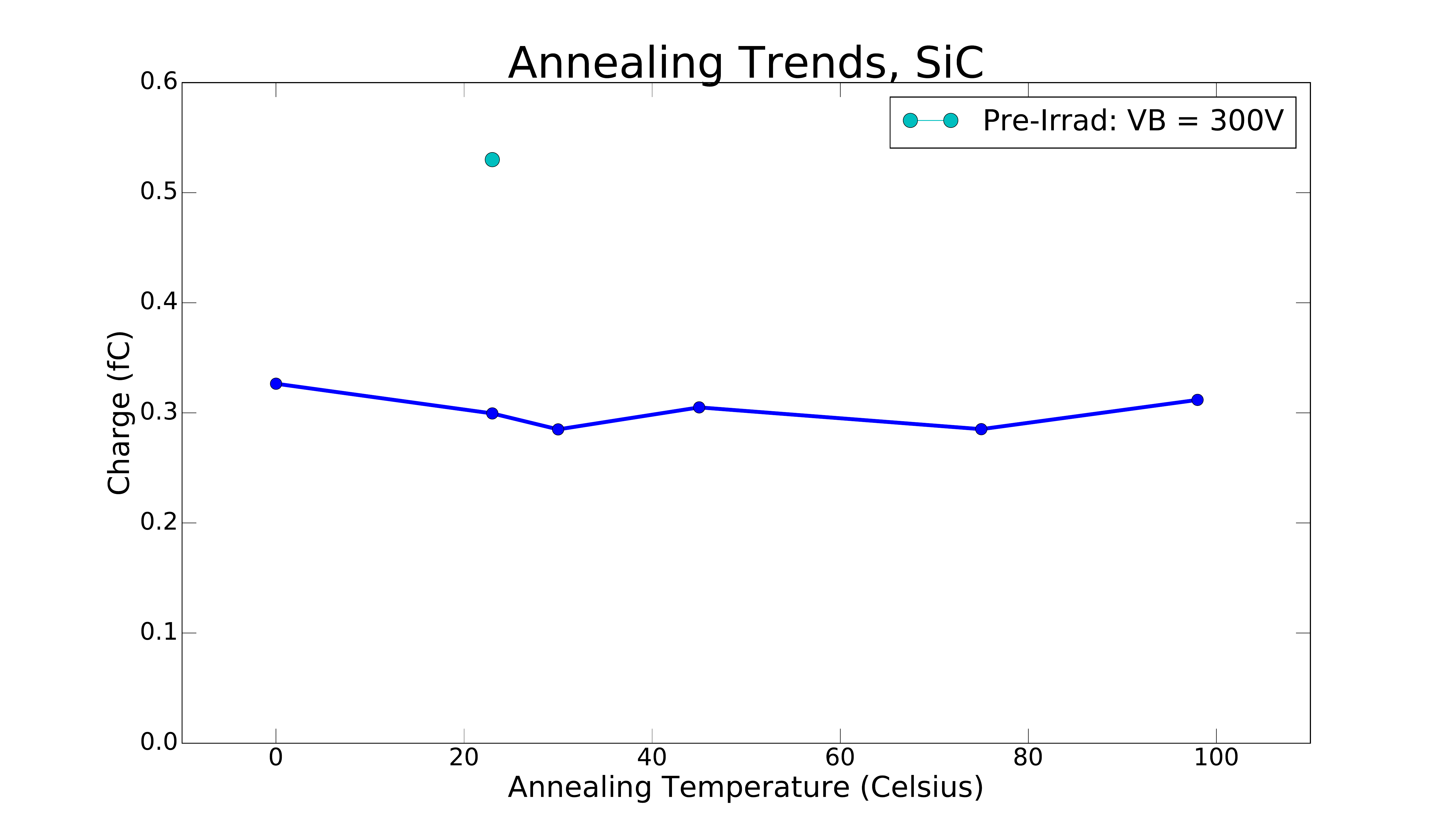}
 \end{center}
 \caption{
Dependence of the median collected charge from a SiC sensor upon
annealing temperature for a bias of $V_B$ = 600 V, before and after exposure to a dose of 77 Mrad
of electromagnetically-induced radiation.
\label{fig:SiC_slice}
}
\end{figure}

As illustrated in Figure~\ref{fig:SiC_traj}, the 
sample sensor of 4H silicon carbide with an epitaxial (active) layer thickness of 
70 $\mu$m exhibited charge collection of approximately 0.5 fC before irradiation. After
an electromagnetically-induced irradiation of 77 Mrad, substantial loss of
CCE was observed at lower bias levels ($V_B \simeq 200$ V); however, the CCE was approximately
2/3 of its unirradiated value for $V_B = 1000$ V. As illustrated in 
Figure~\ref{fig:SiC_slice}, little improvement was observed after hour-long
annealing episodes at successively higher temperature. Post-irradiation
leakage current, measured at approximately $-15^{\circ}$ C, remained at the
nanoamp level.

\subsection{Results for silicon diode sensors}

\begin{figure}[h]
 \begin{center}
   \includegraphics[width=0.50\textwidth]{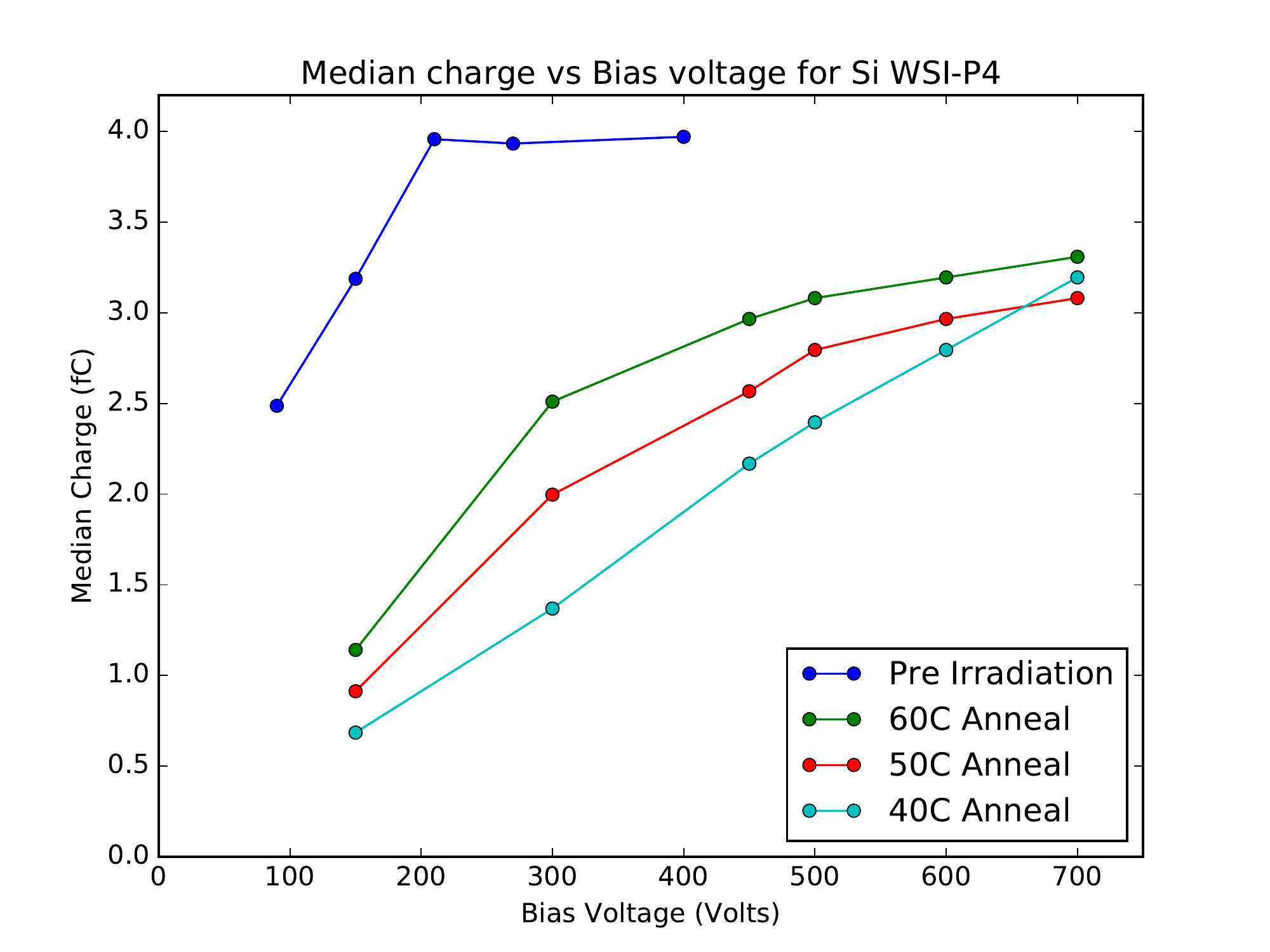}
 \end{center}
 \caption{
Dependence of the median collected charge from a p-bulk
float-zone technology silicon diode sensor upon bias
voltage and annealing temperature, after exposure to a dose of 270 Mrad
of electromagnetically-induced radiation. Also shown is the median
collected charge as a function of bias voltage prior to irradiation.
\label{fig:PF_traj}
}
\end{figure}

\begin{figure}[h]
 \begin{center}
   \includegraphics[width=0.50\textwidth]{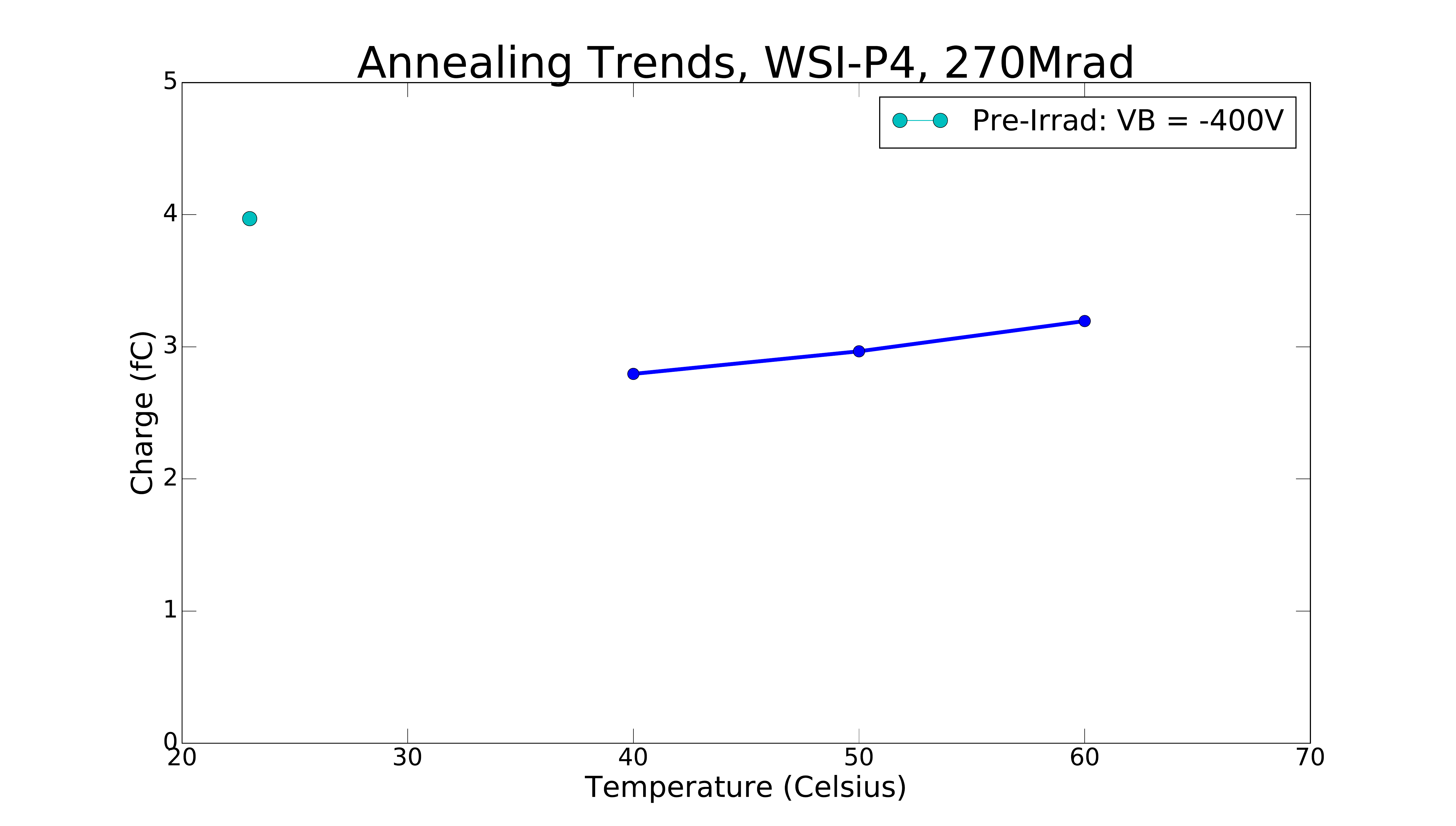}
 \end{center}
 \caption{
Dependence of the median collected charge from a p-bulk, float-zone technology silicon diode sensor
as a function of annealing temperature for a bias of $V_B = -600$ V, before and after exposure to a
dose of 270 Mrad of electromagnetically-induced radiation.
\label{fig:PF_slice}
}
\end{figure}

\begin{figure}[h]
 \begin{center}
   \includegraphics[width=0.45\textwidth]{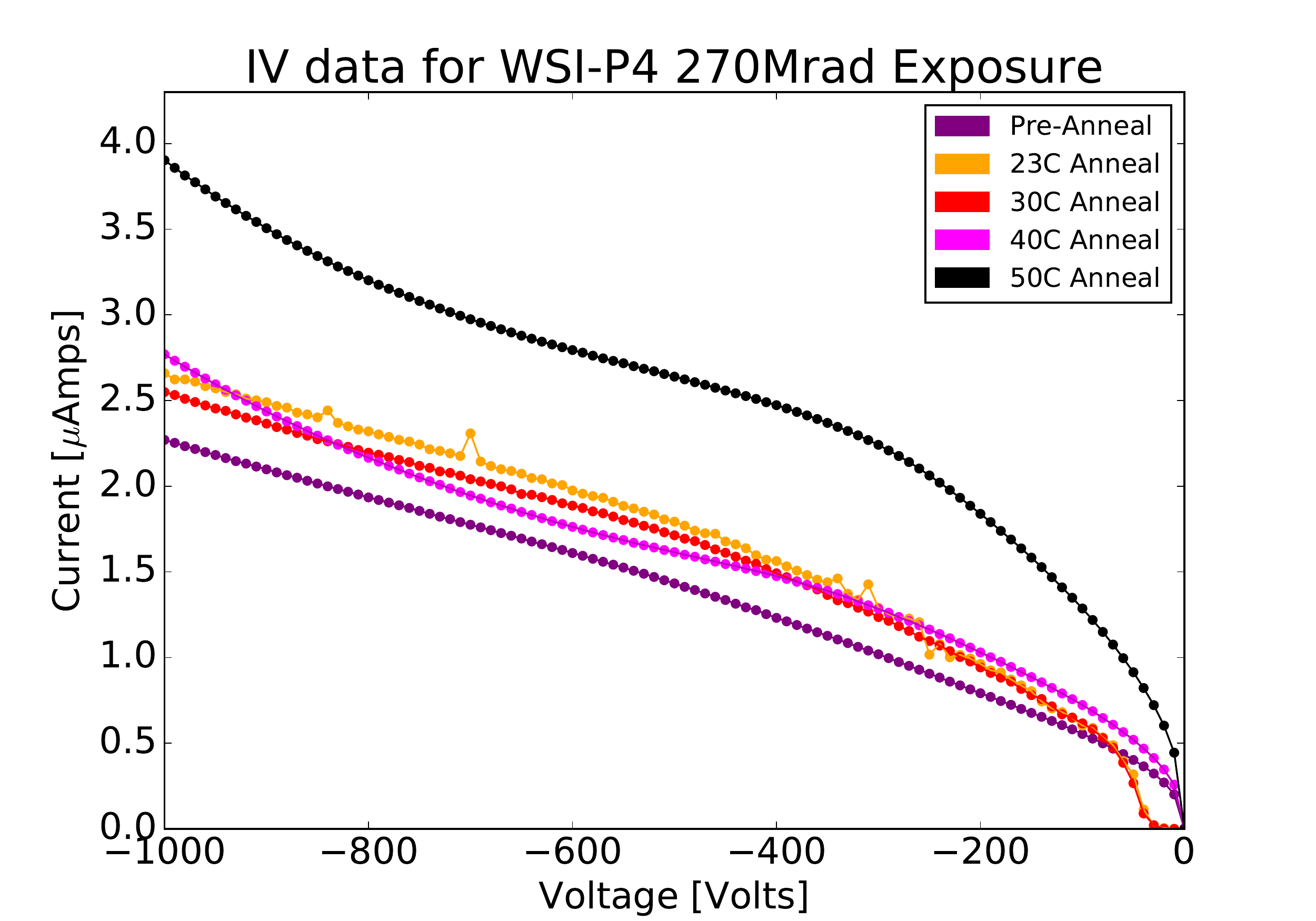}
 \end{center}
 \caption{
Dependence of the leakage current through a p-bulk
float-zone technology silicon diode sensor upon bias
voltage and annealing temperature, after exposure to a dose of 270 Mrad
of electromagnetically-induced radiation.
\label{fig:PF_curr}
}
\end{figure}

\begin{figure}[h]
 \begin{center}
   \includegraphics[width=0.40\textwidth]{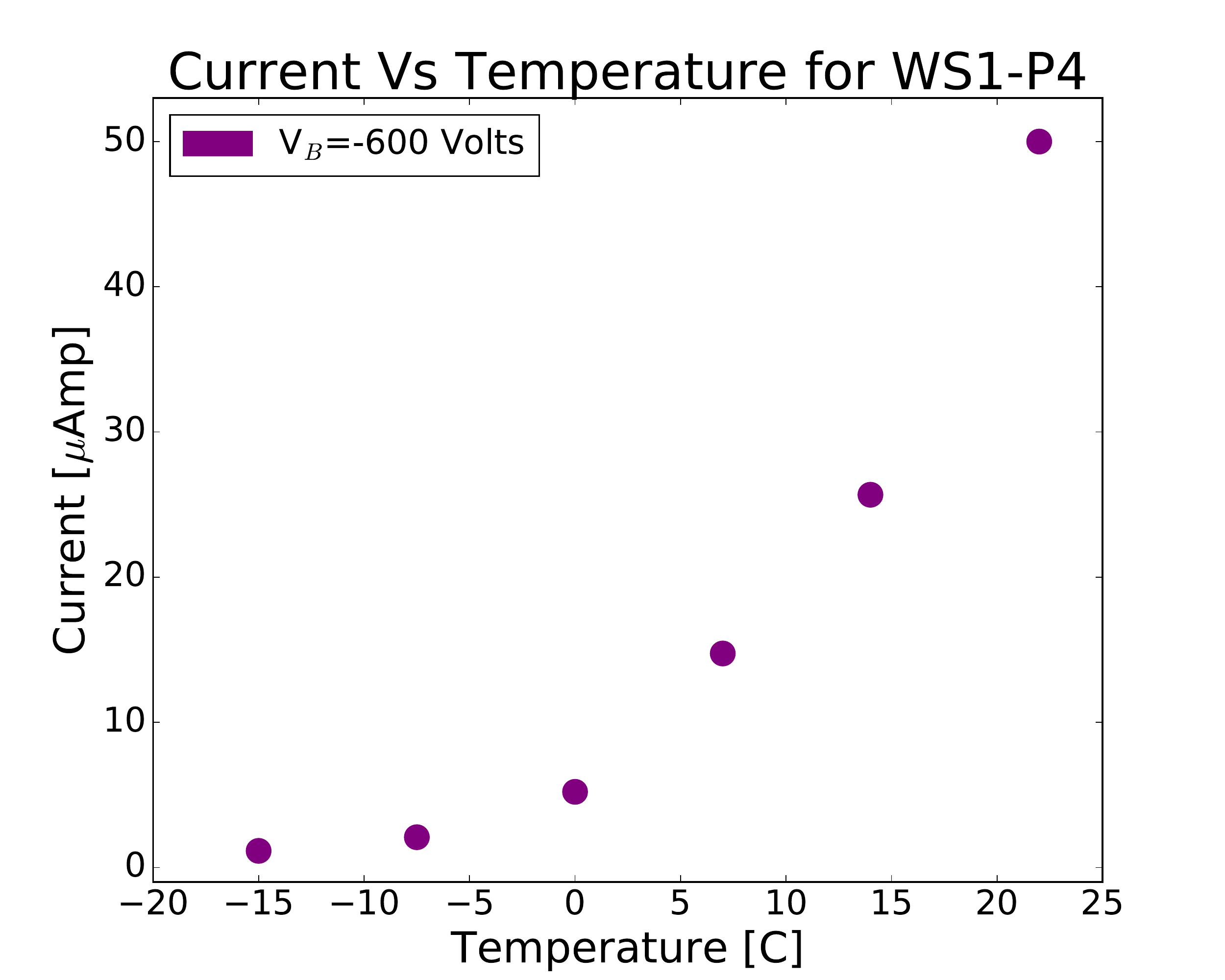}
 \end{center}
 \caption{
Dependence of the leakage current through a p-bulk
float-zone technology silicon diode sensor upon temperature
after exposure to a dose of 270 Mrad
of electromagnetically-induced radiation. The sensor
was biased to $V_B = -600$ V.
\label{fig:PF_curr_T}
}
\end{figure}

CCE and leakage current were measured for three types
of Si diode sensors after doses of electromagnetically-induced
radiation of order 300 Mrad, including both p-bulk (PF) and n-bulk (NF)
float zone pad sensors as well as for an n-bulk sensor (LUMI) designed
for use in the ILC Luminosity Calorimeter.
Figures~\ref{fig:PF_traj} and ~\ref{fig:PF_slice}
exhibit the CCE for the PF sensor
before and after a 270 Mrad irradiation, with the post-irradiation
CCE exhibited after several successive annealing episodes.
Significant CCE loss was observed at lower ($V_B = -200$ V) bias
voltages, but for $V_B = -600$ V, the CCE was observed to exceed 
80\% of its pre-irradiation value after annealing at moderate temperature.
Figure~\ref{fig:PF_curr} shows the PF sensor leakage current
observed after irradiation, measured as a function of bias
voltage at a temperature of $-10^{\circ}$ C. Taking into account
the $0.025$ cm$^2$ area of the sensor, the current
density was measured to be approximately 80 $\mu$A/cm$^2$,
with little dependence upon annealing temperature. Figure~\ref{fig:PF_curr_T}
exhibits the post-irradiation leakage current as a function of temperature
for a bias voltage of $V_B = -600$ V; the power-law behavior
of a doubling of the leakage current for every $5-10^{\circ}$ C
increase in temperature is typical for Si diode sensors.

\begin{figure}[h]
 \begin{center}
   \includegraphics[width=0.50\textwidth]{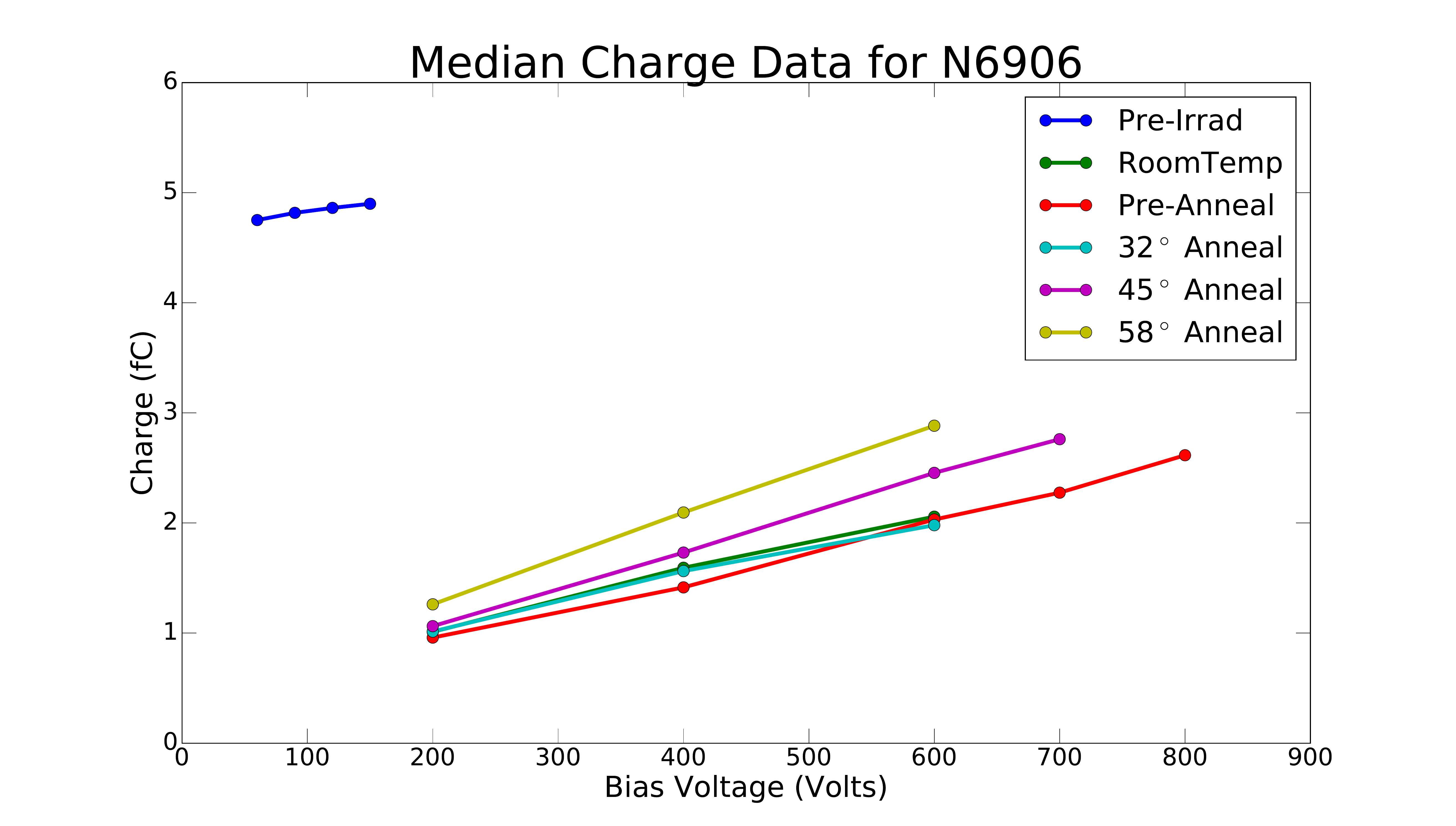}
 \end{center}
 \caption{
Dependence of the median collected charge from an n-bulk
float-zone technology silicon diode sensor upon bias
voltage and annealing temperature, after exposure to a dose of 290 Mrad
of electromagnetically-induced radiation. Also shown is the median
collected charge as a function of bias voltage prior to irradiation.
\label{fig:NF_traj}
}
\end{figure}

\begin{figure}[h]
 \begin{center}
   \includegraphics[width=0.50\textwidth]{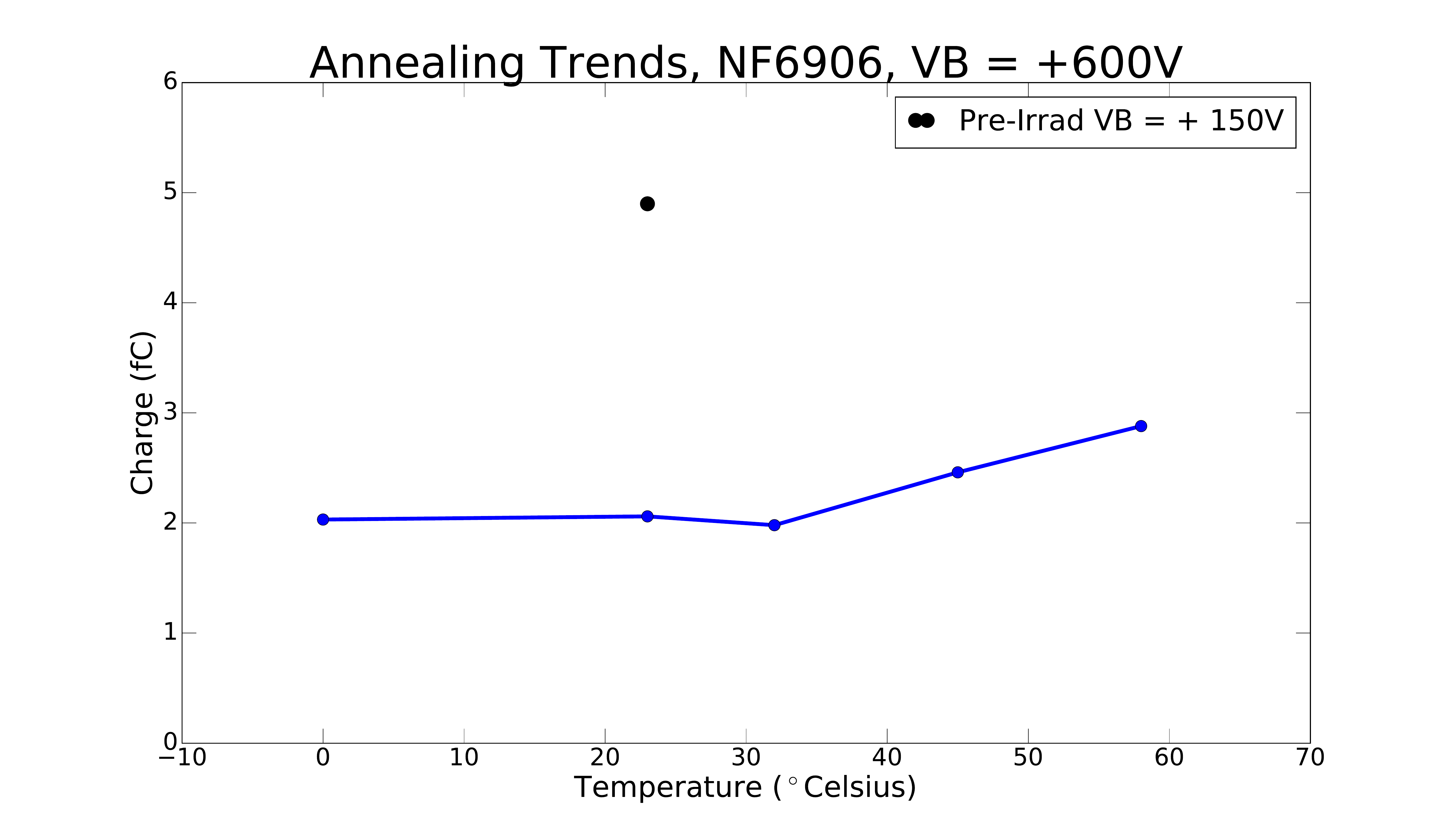}
 \end{center}
 \caption{
Dependence of the median collected charge from an n-bulk, float-zone technology silicon diode sensor
as a function of annealing temperature for a bias of $600$ V, before and after exposure to a 
dose of 290 Mrad of electromagnetically-induced radiation.
\label{fig:NF_slice}
}
\end{figure}

Figures~\ref{fig:NF_traj} 
and~\ref{fig:NF_slice} exhibit the CCE for the NF sensor
before and after a 290 Mrad irradiation, with the post-irradiation
CCE exhibited after several successive annealing episodes.
Again, significant CCE loss was observed at lower ($V_B = 200$ V) bias
voltages, while for $V_B = 600$ V, the CCE was observed to approach
60\% of its pre-irradiation value after annealing at moderate temperature.
The post-irradiation leakage current density was found to be similar in
magnitude and temperature dependence to that of the PF sensor discussed
above.

\begin{figure}[h]
 \begin{center}
   \includegraphics[width=0.43\textwidth]{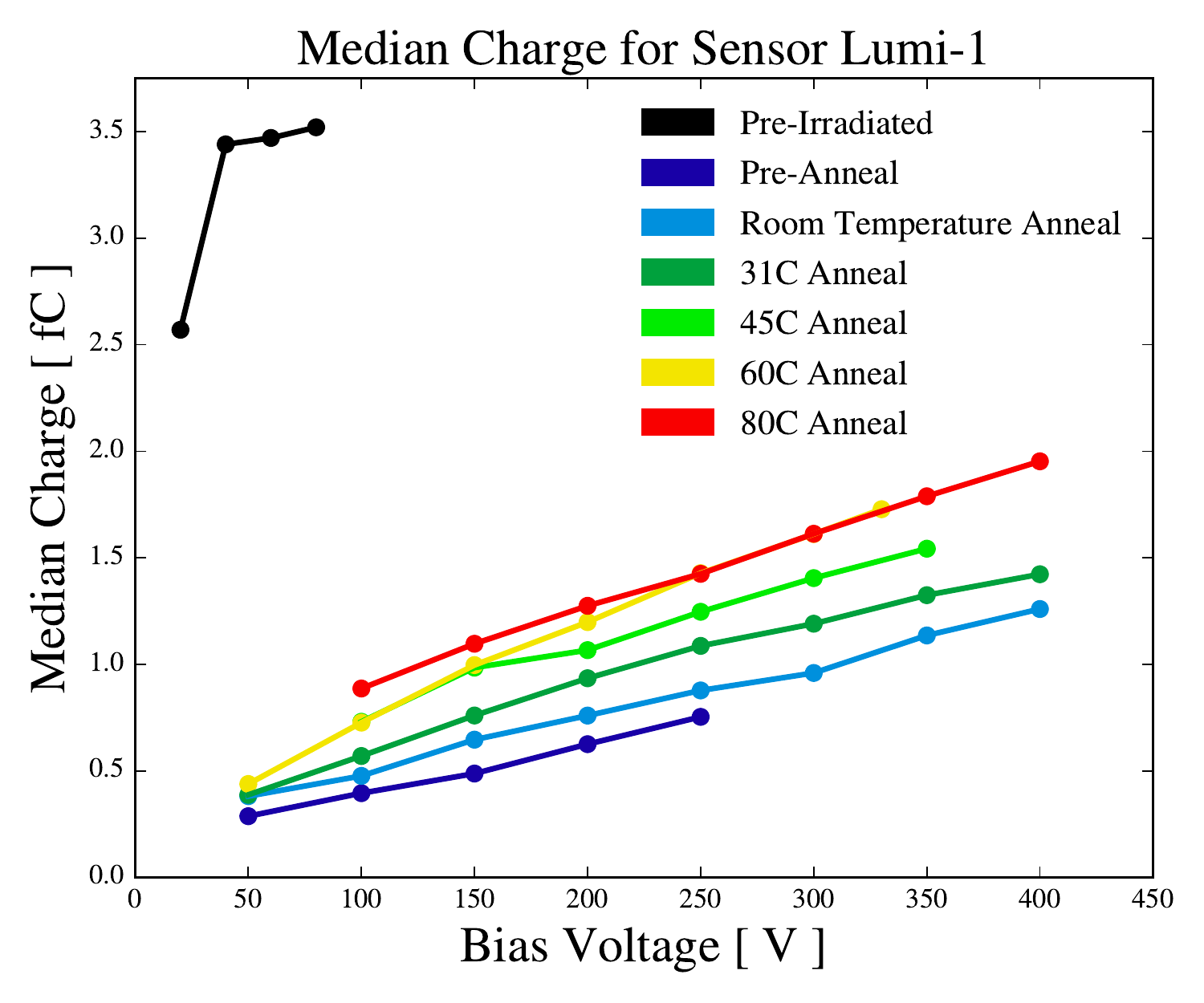}
 \end{center}
 \caption{
Dependence of the median collected charge from a fragment of
an n-bulk silicon diode sensor designed for use with the ILC
Luminosity Calorimeter, as function of bias
voltage and annealing temperature, after exposure to a dose of 316 Mrad
of electromagnetically-induced radiation. Also shown is the median
collected charge as a function of bias voltage prior to irradiation.
\label{fig:Lumi_traj}
}
\end{figure}

Finally, figure~\ref{fig:Lumi_traj}
exhibits the CCE for the LUMI sensor
before and after a 316 Mrad irradiation, with the post-irradiation
CCE again exhibited after several successive annealing episodes.
The results are qualitatively similar to those for the PF and NF
sensors, with significant CCE loss observed for lower bias voltages,
but with substantial improvement to the CCE observed for higher
bias voltages, particularly after annealing at moderate temperature.
Because of the breakdown experienced at bias voltages above several
hundred volts (due, again, to the fact that the integrity of the
sensor was compromised when it was broken and cleaved, but
not expected for an intact sensor), the
high-bias CCE behavior could not be
explored. Nonetheless, the behavior was observed to be qualitatively
consistent with that of the NF sensor discussed above.
Given the possibility that significant current was passing across 
the cleaved edge of the sensor, only an upper bound could be 
determined for the post-irradiation leakage current for the LUMI 
sensor. Again, though, this bound suggested qualitative consistency with the NF
and PF sensor results.

\section{Summary and Conclusions}

Using the End Station Test Beam facility at the SLAC National
Accelerator Laboratory,
we have explored the radiation tolerance of three different types of
silicon diode sensors as well as bulk (non-diode) sensors formed from 
gallium arsenide, silicon carbide, and industrial sapphire crystals.
These sensors were exposed to doses of electromagnetically-induced
radiation that varied from 21 Mrad (for the gallium arsenide sensor)
to of order 300 Mrad (for the silicon diode and sapphire sensors).
At these dose levels, the silicon diode sensors were observed to 
develop leakage currents of several tens of $\mu$A/cm$^2$
at operating temperatures of $-10^{\circ}$C,
increasing with temperature with a doubling interval of $5-10^{\circ}$C.
After moderate-temperature
annealing, and operating at a reverse bias of 400-600 V, 
the silicon diode sensors were observed to retain charge-collection
efficiency above 50\%, with the p-bulk sensor retaining
somewhat better charge-collection efficiency than the two n-bulk
sensors that were irradiated. The silicon carbide and sapphire
sensors, after absorbing doses of 77 and 307 Mrad, respectively,
did not develop significant leakage current at an operating temperature
of $-10^{\circ}$C. Operating at a bias voltage of $V_B$ = 1000 V, the silicon 
carbide sensor retained somewhat over 50\% of its unirradiated charge-collection
efficiency, while the sapphire sensor retained only about 20\% of its
unirradiated charge-collection efficiency. Neither sensors showed
improvement with annealing. Finally, after moderate temperature annealing, 
the gallium arsenide sensor retained approximately 50\% of its
original charge-collection efficiency after a 21 Mrad dose. Somewhat
unexpectedly, the sensor was observed to develop a leakage current
of approximately 1 $\mu$A/cm$^2$ at an operating temperature of $-10^{\circ}$C,
with a fractional rate of increase with temperature similar to that of the silicon
diode sensors. These gallium arsenide results await confirmation with a 
sensor that has been irradiated to 100 Mrad but has not yet been characterized.

\section{Acknowledgments}

We are grateful to Leszek Zawiejski, INP, Krakow for supplying us with the tungsten plates
needed to form our radiator, Georgy Shelkov, JINR, Dubna for supplying us with GaAs
sensors for irradiation studies, 
Bohumir Zatko, Slovak Academy of Sciences, for supplying us with the SiC sensor,
and Sergej Schuwalow, DESY Zeuthen, for supplying us with the industrial sapphire sensor.  
We would also like to express our gratitude
to the SLAC Laboratory, and particularly the End Station Test Beam delivery
and support personnel, who made the run possible and successful.
Finally, we would like to thank our SCIPP colleague Hartmut Sadrozinski for
the numerous helpful discussions and guidance he provided us.

\section{Role of the Funding Source}

The work described in this article was supported by the United States Department of Energy,
DOE contract DE-AC02-7600515 (SLAC) and grant DE-FG02-04ER41286 (UCSC/SCIPP). The funding agency
played no role in the design, execution, interpretation, or
documentation of the work described herein.




\nocite{*}
\bibliographystyle{elsarticle-num}
\bibliography{martin}



\end{document}